\documentclass{sigchi}

\CopyrightYear{2020}
\setcopyright{acmlicensed}
\doi{https://doi.org/10.1145/3313831.3376246}
\isbn{978-1-4503-6708-0/20/04}
\conferenceinfo{CHI'20,}{April  25--30, 2020, Honolulu, HI, USA}
\acmPrice{\$15.00}
\toappear{\scriptsize Permission to make digital or hard copies of all or part of this work for personal or classroom use is granted without fee provided that copies are not made or distributed for profit or commercial advantage and that copies bear this notice and the full citation on the first page. Copyrights for components of this work owned by others than ACM must be honored. Abstracting with credit is permitted. To copy otherwise, or republish, to post on servers or to redistribute to lists, requires prior specific permission and/or a fee. Request permissions from permissions@acm.org. \\
{\emph{CHI '20, April 25--30, 2020, Honolulu, HI, USA.} } \\
Copyright is held by the owner/author(s). Publication rights licensed to ACM. \\
ACM ISBN 978-1-4503-6708-0/20/04\ ...\$15.00.\\
http://dx.doi.org/10.1145/3313831.3376246}

\usepackage{balance}       
\usepackage{graphics}      
\usepackage[T1]{fontenc}   
\usepackage{txfonts}
\usepackage{mathptmx}
\usepackage[pdflang={en-US},pdftex]{hyperref}
\usepackage{color}
\usepackage{booktabs}
\usepackage{textcomp}
\usepackage{subcaption}

\usepackage{microtype}        
\usepackage{ccicons}          

\usepackage{todonotes}

\def\plaintitle{Sociality and Skill Sharing in the Garden}
\def\plainauthor{First Author, Second Author, Third Author,
  Fourth Author, Fifth Author, Sixth Author}

\def\plainkeywords{Gardening; Sociality; skill sharing; Participant Observation}

\makeatletter
\def\url@leostyle{%
  \@ifundefined{selectfont}{
    \def\UrlFont{\sf}
  }{
    \def\UrlFont{\small\bf\ttfamily}
  }}
\makeatother
\def\pprw{8.5in}
\def\pprh{11in}

\definecolor{linkColor}{RGB}{6,125,233}
\hypersetup{%
  pdftitle={\plaintitle},
  pdfauthor={\plainauthor},
  pdfkeywords={\plainkeywords},
  pdfdisplaydoctitle=true, 
  bookmarksnumbered,
  pdfstartview={FitH},
  colorlinks,
  citecolor=black,
  filecolor=black,
  linkcolor=black,
  urlcolor=linkColor,
  breaklinks=true,
  hypertexnames=false
}

\begin{document}

\title{\plaintitle}

\numberofauthors{2}
\author{%
  \alignauthor{Hanuma Teja Maddali\\
    \affaddr{Department of Computer Science\\University of Maryland}\\
    \affaddr{College Park, USA}\\
    \email{hmaddali@umd.edu}}\\
  \alignauthor{Amanda Lazar\\
    \affaddr{College of Information Studies\\ University of Maryland}\\
    \affaddr{College Park, USA}\\
    \email{lazar@umd.edu}}\\
}

\maketitle

\begin{abstract}
Gardening is an activity that involves a number of dimensions of increasing interest to HCI and CSCW researchers, including recreation, sustainability, and engagement with nature. This paper considers the garden setting in order to understand the role that collaborative and social computing technologies might play for practitioners engaging in outdoor skilled activities. We conducted participant observations with nine experienced gardeners aged 22-71 years. Through this process, we find that gardeners continuously configure their environments to accommodate their preferences for sociality. They share embodied skills and help others attune to sensory information in person, but also influence learning through the features in their garden that are observed by others. This paper provides an understanding of sociality in the garden, highlights skill sharing as a key domain for design in this space, and contributes design considerations for collaborative technologies in outdoor settings.
\end{abstract}




\section{Introduction}
HCI researchers are examining outdoor activities and nature-spaces as sites of recreation, learning, and social-interaction. Gardening is one such outdoor activity with ties to other topics of interest in HCI such as food sustainability and civic engagement with environmental issues. Gardening is an activity that builds community and increases residents' attachment to their neighborhood \cite{Litt2015}. It fosters interactions between people from diverse backgrounds and serves as a site for the experiential learning of social and civic skills such as leadership, community organizing, and cultural competency \cite{Litt2015,Twiss2003}. 

People are also drawn to gardening for opportunities to spend time outdoors and with family \cite{nga:5YearYeport}.
A 2014 report estimates that one in three households in the United States engage in food gardening, or urban agriculture \cite{nga:5YearYeport}. These 42 million households include people of all ages. Yet, not all groups have equal access to getting engaged in gardening. Access to land for gardening in an urban setting is limited \cite{armar2000urban}, a lack of gardening knowledge can limit participation \cite{Davis2017}, and for some groups with mobility constraints, technology typically focuses on indoor living rather than outdoor spaces \cite{ Zhang:2017, ZSCHIPPIG2016186}. With a growing interest in the opportunities it presents for technological applications for group interaction, and in the work still needed to promote access to this activity, gardening appears to be an area ripe for HCI research.

In considering design in the garden space, however, it is key to turn to the body of past research that has found that practitioners are sensitive to how technologies are introduced the garden space. For example, automation using sensor-networks can be perceived as obstructing the sensory, embodied, emotional feeling of engaging directly with nature \cite{Heitlinger:2013:SHG:2541016.2541023,Odom:2010:MWD:1858171.1858211, Goodman:2011:GGN:1978942.1979273}. Gardeners may trust their own localized, developed knowledge over scientific models \cite{Pearce:2009:SEU:1738826.1738861}. Poorly designed technology can also impede important social practices in the garden, for example, the transmission of skills from experienced to novice gardeners \cite{Odom:2010:MWD:1858171.1858211}. The drawbacks of purely technological approaches when engaging gardeners, paired with the social nature of gardening, point towards exploring social-computing design approaches as a fruitful area of research \cite{Baumer:2011:ID:1978942.1979275, Lyle:2015:GFC:2768545.2768549}. Understanding the potential role of social technologies in this space requires an examination of where sociality exists in the garden, as well as the particular kinds of interactions that might be supported.

Our research takes, as a starting point, findings from past work: that gardening is sensory and emotional, with social practices that have been built around these activities over time. Given that past work has typically sought gardeners' perspectives on technologies that transmit information to gardeners (e.g., soil quality \cite{Kuznetsov:2011:NNS:2030112.2030144, Kuznetsov:2013:CEL:2466627.2466638} and temperature \cite{Baumer:2011:ID:1978942.1979275} sensors), we return to the garden setting with an ethnographically informed approach, engaging in participant observations with nine gardeners to identify opportunities for HCI. Our paper offers three contributions. First, we provide an understanding of sociality in terms of where it exists in the garden and how gardeners configure desired levels of social interaction, for example, through physical arrangements such as letting vines grow over a fence to obscure the view of those passing by. Second, we highlight skill sharing as a key domain for social design in this space. We find that this process of skill sharing is supported through different levels of engagement with practitioners: in addition to direct interaction and the use of digital platforms, practitioners learn techniques and other information through observation of others’ gardens. Finally, we contribute design considerations for collaborative technologies in outdoor settings with implications for embodied skill sharing and inclusion. We suggest that technology designed for social computing in the garden should balance respecting dynamic sociality preferences with motivating community engagement and collaboration.

\section{Related Work}

Below, we discuss research on technology for outdoor activities in nature-spaces, the garden as a site for community engagement, and perspectives on technology in the garden. 
\subsection{HCI in the Outdoors}
Recent HCI research situates technology design in a range of outdoor activities. These include recreational activities (e.g. tourism \cite{Potter:2016:VRN:3010915.3011854}, scuba-diving \cite{Jain:2016:ITS:2851581.2892503}, paragliding \cite{Kang:2018:FSC:3173574.3174206}), fitness (e.g. running \cite{Mauriello:2014:SFF:2611247.2557299}, wall-climbing \cite{Kajastila:2016:ACW:2858036.2858450}, cycling \cite{Walmink:2014:DHR:2540930.2540970}), and nature activities (e.g. hiking \cite{Posti:2014:UJH:2598510.2598592},  foraging \cite{Liu:2018:DCS:3173574.3173614}, and monitoring wildlife \cite{Cappadonna:2016:CWE:2908805.2909413}). Design in this space has considered motivations to engage outdoors as well as the value that technology brings to personal and shared experiences. Some research on outdoor activities draws on social facilitation theory \cite{ Bond:1982} to support social interaction in fitness groups through revealing information such as the speed of runners \cite{Mauriello:2014:SFF:2611247.2557299} and heart-rate of cyclists \cite{Walmink:2014:DHR:2540930.2540970} to others. These studies discuss insights from in-situ presentation of individual and group performance metrics and its potential for supporting group togetherness and motivations for fitness.

In contrast to work that supports social experiences around outdoor activities, other research helps people disconnect from others in order to become more immersed in nature. For example, research has supported purposeful solitude in nature by informing individuals of nearby hikers \cite{Posti:2014:UJH:2598510.2598592}. With the rising interest of HCI in the outdoors, it is important to understand what design considerations exist for supporting sociality in outdoor spaces. In our work, we find the importance of acknowledging varying, rather than static social preferences in outdoor spaces --- a concept that has previously only been considered in a traditional indoor office setting. We discuss how designing for learning or skill sharing in the garden space should account for these varying social preferences.

Nature-spaces are another domain of interest in outdoor HCI, including research that explores the role of technology to support collaboration. Research has examined the design of collaborative technology for search and rescue teams trying to maintain situational awareness in wilderness \cite{Jones:2018:DDC:3272973.3272978} and to support simulations in high-risk outdoor recreation areas \cite{Desjardins:2016:ABP:2818048.2835200}. Nature-spaces also provide an avenue for environmental learning. For example, Soro et al. describe an IoT "Ambient Birdhouse" designed to interest children in engaging with nature by becoming more aware of bird calls and discuss how it could be used as a catalyst for learning and socializing \cite{Soro:2018:ABI:3173574.3173971}. Liu et al. designed three wearables for mushroom foraging in order to "offer a vision of wearables extending our human sensory capacities into the environment" \cite{Liu:2018:DCS:3173574.3173614}. This vision offers people the capacity to "notice, attend to, and become struck by nonhuman lives" and, in the case of one of the prototypes, also share the information that they gather with others \cite{Liu:2018:DCS:3173574.3173614}.

\subsection{Community engagement in the garden} 
A common approach in sustainability HCI research is to design for community engagement to encourage strong civic activity around pro-environmental goals. For example, YardMap supports professionals and citizen-scientists in mapping personal carbon-neutral yard practices, learning about their local environment, and discussing their potential impact on habitats \cite{Dickinson2014}. The inclusion of people from different cultural backgrounds \cite{Heitlinger:2018:CSS:3210604.3210620} and expertise levels \cite{Dickinson2014} through knowledge-sharing \cite{Lyle:2015:GFC:2768545.2768549} and capacity-building \cite{ Heitlinger:2013:SHG:2541016.2541023} is seen as an important mechanism for building sustainable communities. 

Jrene Rahm's work, centered around an inner-city youth gardening program, highlights the role of active social participation in creating opportunities for developing expertise as a novice \cite{rahm2002emergent}. Novices, through situated learning, gain skills and an understanding of the community culture through their interactions with peers or more experienced practitioners and immersion in the garden environment \cite{rahm2002emergent, lave_wenger_1991}. The ways that gardeners become experts in using their senses to notice and observe lead to opportunities to support people in new ways of engaging with the world and with other practitioners towards more sustainable futures \cite{Goodman:2011:GGN:1978942.1979273,Odom:2010:MWD:1858171.1858211,Kuznetsov:2016:EFS:2858036.2858363}. 
However, researchers have also noted the tensions that arise when designing for engagement in communities with diverse expertise levels. For example, managing the territorial behaviors of experts is important in encouraging participation from novices and allow them to develop a feeling of attachment and ownership towards the community \cite{Thom-Santelli:2010:YKE:1753326.1753578}. In our paper, we acknowledge the value of garden spaces in cultivating a sense of community and engaging with other practitioners for skill sharing. We discuss this in light of the tensions that we find in encouraging community inclusion while maintaining ownership of the garden space.

\subsection{Gardeners' Perspectives on Technology}
Several studies have focused on gardeners in community and residential settings. This past research has found that gardeners do use a number of digital tools, often to share information or support coordination. For example, Wang et al. analyze collaboration between gardeners, finding that they use different tools for information and knowledge sharing as well as scheduling work activities \cite{Wang:2015:ISS:2768545.2768556}. In a study of handwork, Goodman and Rosner note that though gardeners and knitters use many different digital tools, they define their own values in opposition to stated negative characteristics of technology \cite{Goodman:2011:GGN:1978942.1979273}. For example, being \textit{engaged} rather than \textit{disconnected} means "committing to the material details of making objects oneself," rather than using a system to cut oneself free of the task of watering \cite{Goodman:2011:GGN:1978942.1979273}. 

A common theme of the body of work on gardening is these tensions that arise with technology. For example, Lyle et al. highlight how gardeners learn through experimentation and observation, as well as the importance of sharing knowledge between community members \cite{Lyle:2015:GFC:2768545.2768549}. In this context, Baumer and Silberman present sensor nets for data-driven gardening as a case study of when the implication is \textit{not} to design a particular technological solution \cite{Baumer:2011:ID:1978942.1979275}. Sensing systems to support automation of tasks or decisions (e.g. automatic watering based on soil moisture) appear to be viewed negatively by gardeners in much past research, as they are seen as interrupting existing values and processes, such as direct interaction between gardeners and plants \cite{Heitlinger:2013:SHG:2541016.2541023,Odom:2010:MWD:1858171.1858211} that help gardeners develop environmental knowledge and intuition \cite{Odom:2010:MWD:1858171.1858211}. Further, automation can interrupt the transfer of knowledge between senior and novice members of the community \cite{Odom:2010:MWD:1858171.1858211}. 

This past literature lays the groundwork for investigating collaborative technologies and social computing in community and residential gardening as a fruitful area of investigation, in that it matches the social and collaborative nature of gardening and moves away from purely technological solutions \cite{ Baumer:2011:ID:1978942.1979275}. Understanding the potential role of collaborative technologies requires filling a gap in our understanding of how sociality manifests in the garden. In our paper, we discuss our finding on two such components of sociality: how a practitioner’s sociality preferences are reflected in their working space, as well as the level of active engagement with other practitioners when teaching and learning skilled activities in the garden.

\section{Methods}
 Fieldwork was conducted over the summer (June through September 2018) in the Mid-Atlantic region of the United States. Below, we describe study procedures, participants who took part in this research, and our analysis.

\subsection{Study Procedures}
We took an ethnographically informed approach to data collection and analysis \cite{Emerson:2011}. Sessions involved a 15-minute interview, brief drawing prompt, and 60-minute participant observation session. The interview included questions such as participants' motivation, frequency of gardening, and self-described level of expertise. In the drawing prompt, participants drew the physical sites where they gardened, including the places that held meaning as well as where they grew different plants and kept tools. In the participant observation, gardeners were asked to engage in activities that they normally would do around their garden. The first author shadowed and worked alongside gardeners and asked questions when relevant to the task at hand, including how gardeners made particular decisions. 

Data collected included the sheet from the drawing prompt, observation notes, video, and audio recordings. Video was collected with a head-mounted GoPro Hero 5 action camera. Our sessions yielded approximately 800 minutes of audio and video recordings (93 minutes per session on average). In parallel with data collection, both authors spent time becoming familiar with the process of gardening. The first author participated in weekly volunteer sessions at the university community garden for four months, and the second author engaged in gardening in her backyard. These experiences informed the study protocol and our understanding of gardening practices and the process of learning gardening skills.

\subsection{Participants}
\begin{table}
  \centering
  \begin{tabular}{lllll}
    \toprule
    \small{\textit{ID}} & \small{\textit{Age}} & \small{\textit{Gender}} & \small{\textit{Ethnicity}} & \small{\textit{Observation Site}}\\
    \midrule
    P1 & 60 & Male & - & Public Community Garden\\
    P2 & 26 & Female & White & Home Garden\\
    P3 & 61 & Female & White & Home Garden\\
    P4 & 37 & Male & Indian & University Community Garden\\
    P5 & - & - & - & Private Community Garden\\
    P6 & - & - & - & University Community Garden\\
    P7 & 37 & Female & White & University Community Garden\\
    P8 & 71 & Male & African & Public Community Garden\\
    P9 & 22 & Female & White & Shade Garden (Ornamental)\\
    \bottomrule
  \end{tabular}
  \caption{Self-Reported Participant Information ("-" indicates participant wished to keep information private)}~\label{tab:participant_info}
\end{table}

Nine participants who self-identified as gardening regularly were recruited through local community garden e-mail lists, fliers posted on campus, word of mouth, and snowball sampling. Nine individuals between the ages of 22 and 71 ( average=45 years, std. dev=19.3 years) participated in the study. All sessions involved participant observations, but four individuals did not engage in the initial 15-minute interview (P5, P6, P7, P9) and two did not engage in the drawing session (P1, P6) due to time constraints.

We attempted to recruit from a range of gardening configurations in order to understand how experiences may vary in different spaces. Participants gardened in different arrangements, from private backyards to public community gardens\footnote{A community garden is a single piece of land gardened collectively by a group of people. Table 1 indicates whether these gardens were situated on private or public land. Public community gardens are usually managed by local government.} (Table \ref{tab:participant_info}). All participants grew food items, the most common being tomatoes, peppers, and herbs. Almost half of the participants grew flowers for themselves, and the majority grew flowers for pollinators. During the participant observation, we asked individuals to engage in whatever tasks they might naturally be doing that day. This ended up including a variety of activities: weeding, watering, trellising, harvesting, decorating, and just relaxing in the garden.

\subsection{Analysis}
Our constructivist grounded theory approach to analysis \cite{Charmaz2014} was as follows: the first author open-coded two transcribed interviews and three sets of observation notes to create a preliminary set of codes and emerging themes. The research team met to discuss these codes over several sessions and became interested in themes relating to \textit{Sociality} (with codes such as "being accessible to passersby," "having informal boundaries in shared plot," and "viewing the garden from an outsider's point-of-view") and \textit{Skill Sharing} (with codes such as "learning from someone who seems more experienced," "observing decorations on neighbors plot," and "sharing a photograph to describe plant condition"). The first author then coded the rest of the transcribed interviews for these themes, adding additional codes as they emerged. The research team related codes to each other through an iterative process of memoing and theorizing, engaging in constant comparison of data to understand and refine a set of high-level themes. 

\begin{figure*}[ht]
  \begin{subfigure}[b]{0.49\linewidth}
    \includegraphics[width=\linewidth]{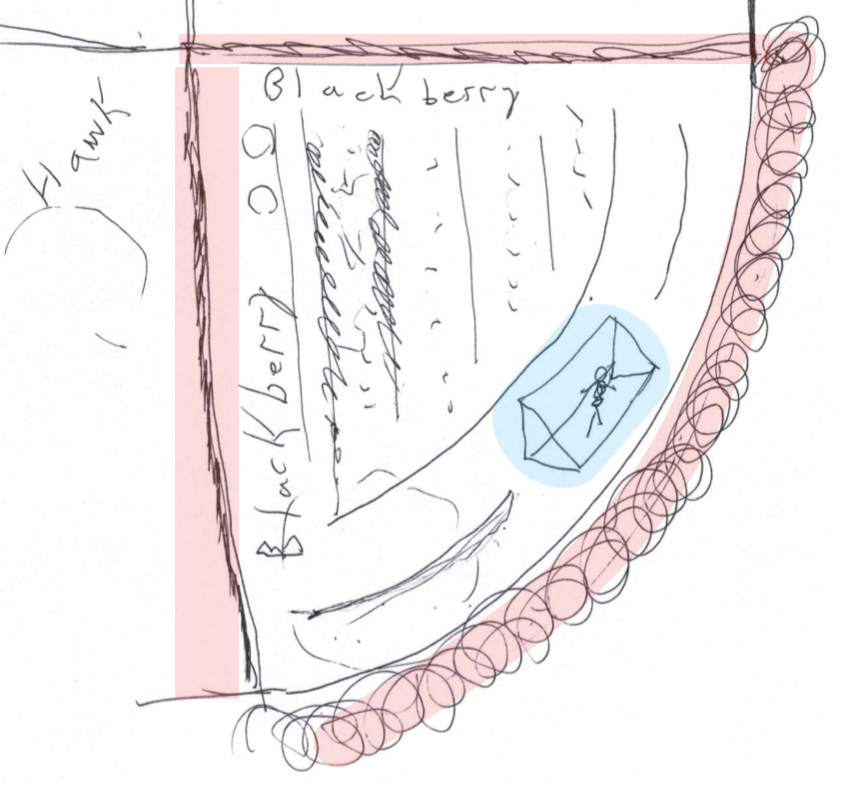}
  \end{subfigure} \begin{subfigure}[b]{0.49\linewidth}
    \includegraphics[width=\linewidth]{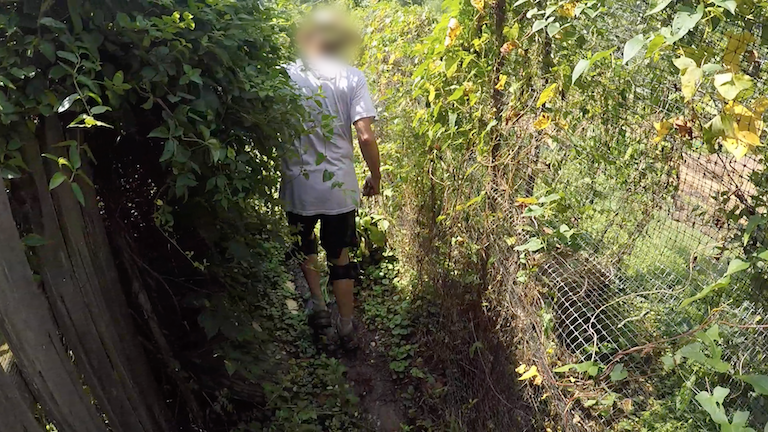}
    \includegraphics[width=\linewidth]{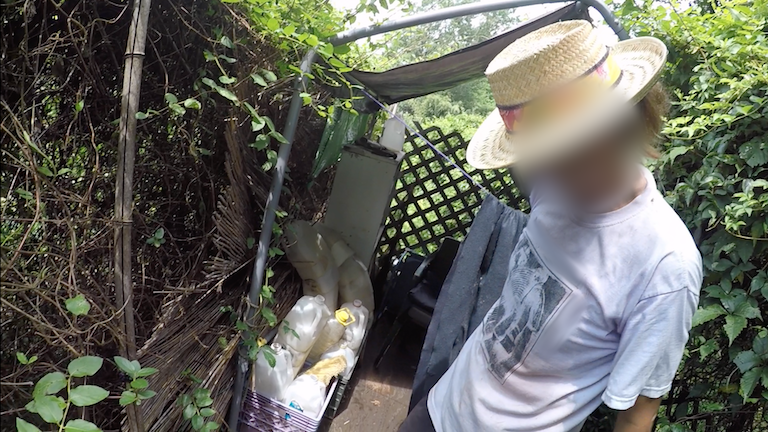}
  \end{subfigure}
  \caption{P5's drawing of his "secret garden" (left) shows blackberry and honeysuckle covering the fence (highlighted in red) to create a sense of privacy and a meditation hut (highlighted in blue).}
  \label{fig:P5_drawing}
\end{figure*}

\subsection{Limitations}
Though participants were diverse in terms of the range of settings in which they gardened, the small number of participants, their all being based in the US, and our approach to recruitment and analysis means that these findings are not intended to be generalizable. The emphasis on skill sharing that arose was likely shaped by our method of participant observation and participant roles in gardens (e.g., working in a community learning garden). Future work is needed to examine a more diverse and comprehensive sample. 

\section{Findings}\label{findings}
In this section, we discuss where sociality arises in the garden and the ways gardening skills are taught and learned. We find that gardeners configure desired levels of sociality. One way they do so is through physical arrangements. Elements such as the type of fencing (e.g., honeysuckle-covered fences or a chain link fence) can indicate ownership and also manage interaction with other gardeners and passersby. One kind of interaction that takes place in the garden is the learning and teaching of skills between gardeners. Skills are shared directly, and also indirectly through observation of others' gardens.

\subsection{Configuring sociality}
Participants engaged in different levels of sociality. This appeared in the actual ways they went about gardening: P5 intentionally gardened alone, P2 worked with her partner on most major tasks, and P4 liked to engage with and learn from others in his community garden. P3 touched on the ways that different gardeners might be drawn to different kinds of gardening arrangements: where she gardened, \textit{"it's close, you share a lot of space... people that want to be on their own, they wouldn't come here."} Many gardeners, though, were not solely social or private gardeners -- they chose to be private or social depending on their mood or the activity they were doing.

\subsubsection{Managing interactions with other gardeners}
Gardeners use physical features of gardens, such as raised beds and hedges, to support desired types of social interactions. P1 explained that where he gardened, plots involved \textit{"raised beds with wood around them so we each know our boundaries."} P5 configured his space by letting vines grow on top of existing separations of plots in his community garden to get more privacy: he pointed to the fence on his plot (Figure \ref{fig:P5_drawing}) \textit{"where all the honeysuckle grows... [the] privacy gives me the secret garden feeling that I like."} He appreciated being alone in the garden to find space for introspection, meditation, and the feeling of getting away from culture. P5's case shows a gardener using natural and built features to create a more private space in a community garden. In a contrasting example, P3 sometimes shared tasks with her neighbors in a backyard that included both their garden spaces. It made a private space more social by bringing together \textit{"people that are okay being close to other people."}

\begin{figure}[h]
\centering
\includegraphics[width=0.75\linewidth]{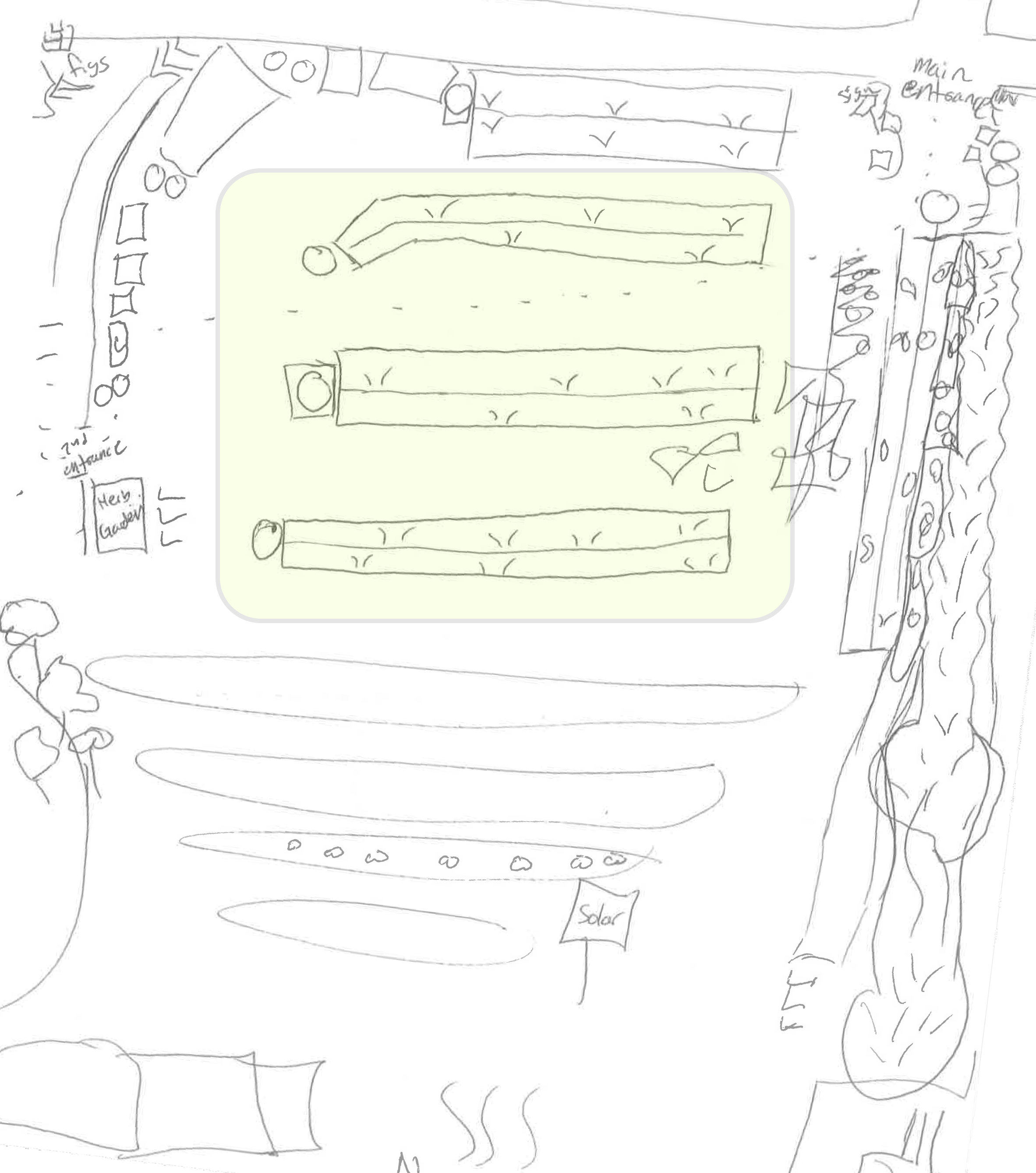}
\caption{Drawing by P9 with ADA accessible teaching spaces (highlighted in green) that are open to walk through for all visitors.}
\label{fig:P9_drawing}
\end{figure}

In P5's example, letting honeysuckle grow wild created a desirable social arrangement for him. P9, on the other hand, spoke about how she arrives at a socially desirable space by picking up debris -- though she leaves leaf litter and smaller sticks as a way of \textit{"keeping it natural"}. P9 clears the pathways in the garden oof debris to make it \textit{"functional for everyone,"} including those with disabilities. She explained that part of the community learning garden where she worked was designed to comply with the Americans with Disabilities act\footnote{The ADA is a civil rights law that prohibits discrimination against individuals with disabilities in all areas of public life, including jobs, schools, transportation, and all public and private places that are open to the general public. For more information, see https://www.dol.gov/general/topic/disability/ada}, \textit{"So people in wheelchairs or with disabilities can easily access this part of the garden, and with the raised beds they can participate in gardening just as well [as] people who don't have a disability"} (Figure \ref{fig:P9_drawing}). P9 appreciated \textit{"how well this space includes a wide range of people."}

Like in the example above, physical configurations (in this case, the removal of large debris) are ways that gardeners reach desired levels of sociality not only based on personal preference, but also based on policy or community-wide decisions. Some community gardeners talked about how garden managers played a role in enforcing community rules of particular sites. Managers resolved issues with a plot that could affect other members, such as directing members to remove weeds that could spread to other plots [P6]. Managers also disseminated news related to group activities, such as putting down wood chips on paths. Even when gardeners were coordinated to support adherence to policies about community space, preferences for privacy could be preserved. P1 explained that \textit{"the manager can send a message to all of us ... we see there's a list of emails, but we don't know which necessarily from the address refers to which person. [We] Certainly don't know which person refers to which plot."} 

\subsubsection{Interaction with those outside the garden}
Participants also configured social interactions not only for fellow gardeners, but also for the broader community that comes into contact with gardens both directly and indirectly. In Heitlinger et al.'s study, a central value of a farm garden is inclusion \cite{Heitlinger:2013:SHG:2541016.2541023}. In our study, we found that a commitment to inclusion shared by many gardeners was in tension with preferences to create divisions between the garden and outside world. Participants saw boundaries that they created or that were features of the space not only as important to keep out animals and people who might take produce or flowers, but also to create a sense of privacy. P3 enjoyed seeing passersby who would complement her flowers, saying that, \textit{"[it] is very nice because you see people coming, passing by ... They are far away enough that they are not in your space..."}. Her garden was separated by an informal boundary created by elevation from the passersby in a shared green space.

\begin{figure}[h]
\centering
\includegraphics[width=\linewidth]{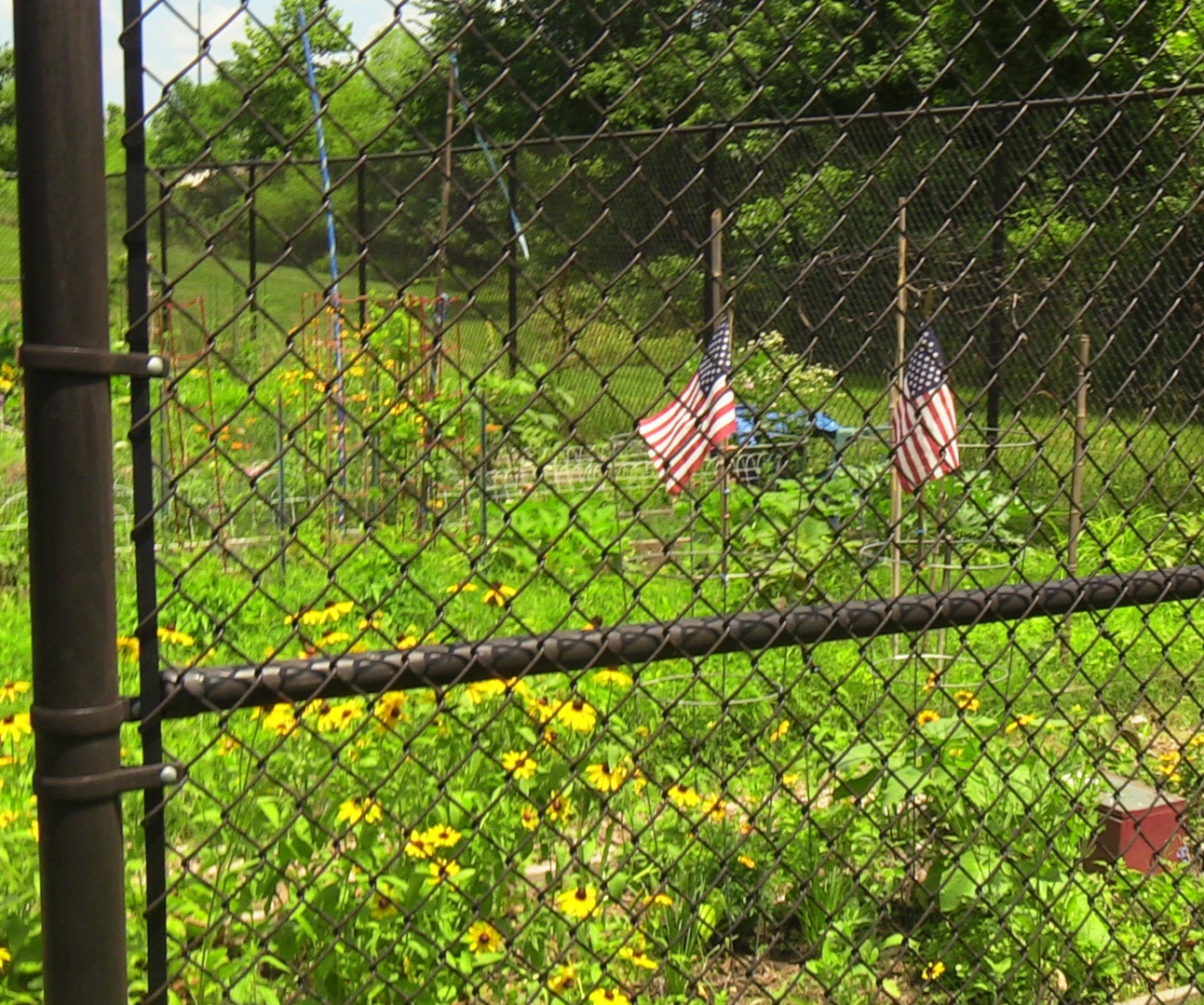}
\caption{P1 decorated the garden for a national holiday with flags, visible through the chain link fence, to show community sentiment.
}
\label{fig:P1_flags}
\end{figure}

Though these separations were important for gardeners to achieve a level of social interaction that was desirable for them, they did think about the ways that some barriers to the outside world might come off as uninviting and worked to create a more welcoming space without necessarily letting others into the garden itself. P1's plot in the community garden had a wire fence with a lock on it. When asked if the garden saw visitors, perhaps children and their parents, from the bordering playground he mentioned that the fence might have unfortunately created a feeling of exclusion for the community: \textit{"There's a sense, perhaps because of the fence, that the gardeners want to be left alone and outsiders don't bother [with them]."} P1 discussed how he decorates his garden, for example with flags for the US holiday July 4th (see Figure \ref{fig:P1_flags}), to show \textit{"community sentiment,"} because, \textit{"people outside the fence and [who] can't get in might feel a little less excluded and maybe it's good public relations for the garden."} P1's chain-link fence allowed individuals to see the decorations he placed in his garden. P8 described an arrangement outside a community garden that encourages community inclusion while preserving boundaries: benches arranged just outside the fence, in a way that invites outsiders to sit and observe the garden.

While some participants like P7 described gardening as an opportunity to \textit{"disconnect from a lot of technology,"} smartphones and social media played a visible role when managing interactions with people outside the garden as well as those in its physical proximity. Participants (P1, P3, P4, P6, P7, P8, P9) frequently mentioned capturing photographs using their smart phones and sharing them, for example, on Facebook and Instagram. These were usually used to enable a passive form of interaction through the posted photographs and other content like, for example, comments that compliment the garden (P1, P4, P8). 

Gardeners also described encouraging active involvement with the garden through volunteering opportunities (P6, P7, P9). P4 describes one such activity where he posted online to invite others to make and share a sauce with peppers. In general, we also find that participants described interactions of a mostly positive nature (P1, P3, P4, P8), both online and offline. P1 explains that its \textit{"a polite thing"} since \textit{"people get sensitive in the garden. Something about the place, something about the activity, you very much want to hear praise."} Though social sharing was an important usage of photographs, photographs were also sometimes taken to keep a personal record of the garden's progress over time (P1, P3, P8). For example, P3 described taking, \textit{"photos of the flowers, because it's fun to remember when they bloom or just they are beautiful."}

\subsubsection{Cultivating desired emotional states and relationships}
In addition to providing or preventing others from access to gardening spaces, gardeners also used physical arrangements to cultivate certain emotional states for themselves. P2 and P3 placed chairs in or within view of their gardens and spent meals and time with partners enjoying the ambiance. Two gardeners described feeling meditative when gardening, with P5 reserving a space for meditating in the garden -- his \textit{"little meditation hut"} (see Figure \ref{fig:P5_drawing}). These findings are consistent with work from anthropology discussing boundaries within the garden itself, where, "there are also separations between different areas and particular functions and activities associated with each" \cite{Alexander:2002}.

Plants and objects in the garden also became ways that participants connected with others outside the garden. Over the course of the study, the first author was offered the following items from participants: beans, tomatoes, eggplants, peppers, strawberries, ground cherries, basil and three types of flowers -- zinnia, globe amaranth, and ageratum. A few gardeners grew items specifically to give as gifts, such as P2 who grew catnip for her friend's cat. Past work has also found that gardeners share produce with friends or fellow gardeners \cite{Goodman:2011:GGN:1978942.1979273, Heitlinger:2013:SHG:2541016.2541023}: we find that gardeners also gave away produce to benefit the community at large. P7 explains, \textit{"we do harvest a lot of stuff and donate it to the campus pantry... because 15 percent of our student population is food insecure."} P4 and P8 shared produce with community members at their place of worship. P4 describes they do this in part because \textit{"it helps save the [place of worship] some money."} 

In addition to connecting to other individuals or a broader community in the present, participants used gardening to reinforce feelings of connections to people or places from the past. P8, who grew hot peppers on one of his plots, referred to how people from his native country love those peppers. P3 described how her hellebore plant \textit{"reminds me of my mom, because she always had them,"} and her gardening toolbox housed a tool that reminded her partner of his father: what she called a \textit{"memory object."} These examples highlight how the garden is shaped to create a personal space that reflects the gardener's relationship with a community or loved one. 

\subsection{Skill sharing in the garden}
A form of interaction that we discuss in this paper is skill sharing in the garden, a recurring theme in  our interviews and observations. More experienced gardeners in our study often had formal and informal teaching roles, but even a master gardener such as P7 acknowledged that \textit{"you never stop learning in this job, which is one of the other reasons why it's so enjoyable."} We detail the different forms in which knowledge and skill sharing about the sensory-rich, embodied practices of gardening took place.

\subsubsection{Tacit knowledge communicated through co-located learning}
In past work, being physically collocated with other gardeners allowed novices to get help from more experienced individuals, for example when dealing with slugs \cite{Wang:2015:ISS:2768545.2768556}. Our findings reveal that co-located gardening enables gardeners to benefit from verbal instruction, but also from non-verbal information and the communication of tacit knowledge. 

\begin{figure}[h!]
\centering
  \begin{subfigure}[b]{\linewidth}
    \includegraphics[width=.9\linewidth]{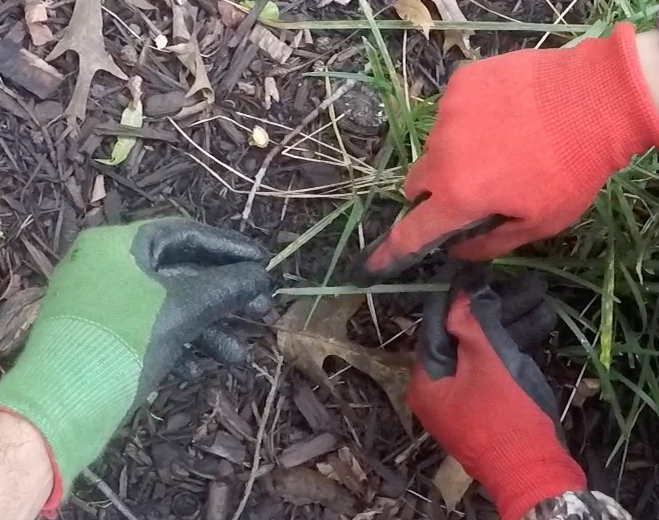}
  \end{subfigure} \begin{subfigure}[b]{\linewidth}
    \includegraphics[width=.9\linewidth]{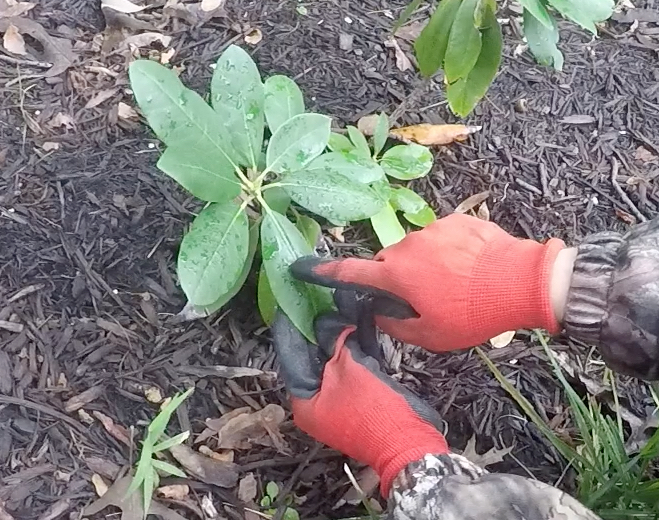}
  \end{subfigure}
  \caption{P9 showing the first author the difference between leaf venation of a monocot (above) and dicot (below).}
  \label{fig:P9_veination}
\end{figure}

Most participants described learning gardening skills from others with more experience in face-to-face interactions. P4 told us that when he was starting out learning to garden, he spent time talking to experienced gardeners and \textit{"picking their brains."} Even now, he enjoyed being in a community gardening setting, because with all the activity in the space, \textit{"I can learn a lot. I feel like it's made me a better gardener."} In our participant observations, we saw the importance of face-to-face sharing to teach embodied and sensory skills. The first author was taught, for example, to measure ripeness using touch by P8 and P2, and to find locations for incisions on the plant and the safe handling of pruning shears by P7. Another anecdote that indicates the importance of face-to-face interaction to communicate knowledge took place when P9 mentioned that it was sometimes, \textit{"hard for me to explain a plant versus a weed... especially if [the people I am teaching] are newer to gardening."} The weeds P9 and the first author were looking for in that spot in the garden were from the dicot plant group. When the first author said that he didn't know what a dicot plant was, P9 showed how the orientation of veins differ between the two: tracing the outline of the leaf veins with her hands, she explained: \textit{"You have this vein here, but then you have these little veins coming off the sides ... so they're not parallel."} This information would have been difficult to communicate without a shared field of view, gestures, and haptic feedback from the veins: all elements that can be seen as inherent to  face-to-face interaction.

As the first author worked in this setting and was taught to notice plants in different ways by experienced gardeners, he began to develop a competence for recognizing different plants based on sensory information. P2 presented a contrast between the leaf texture of pumpkin plants, which the researcher found to be \textit{"crackling"} [excerpt from field notes] and gourds leaves, which P2 explained were \textit{"much softer than the pumpkin."}

\subsubsection{Continuing to learn outside the garden}
\begin{figure}[h]
\centering
\includegraphics[width=0.8\linewidth]{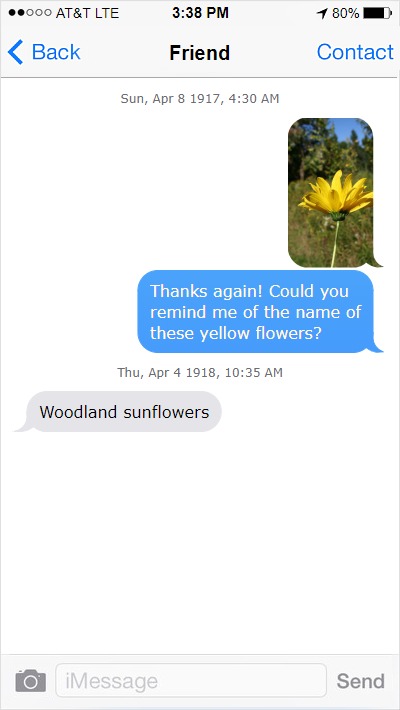}
\caption{Mock-up of mobile screenshot showing how P3 asked for her friend's help with identifying a plant.
}
\label{fig:P3_Message_Friend}
\end{figure}

The affordances of face-to-face skill sharing were clear in our findings. However, as has also been found in past work, gardeners also learned by using a range of digital resources to find information \cite{Goodman:2011:GGN:1978942.1979273}. In our study, this occurred predominantly via text and images, such as how-to blogs (P1), social media messages (P4), and YouTube videos (P2, P5, P8). P4 described sharing an image of a diseased plant with people at a nursery to identify the disease, and P1 told us that he posted pictures of potential weeds online so others could identify them. Individuals also used digital technologies to get information from those they knew: P3 showed us a picture she had messaged to her friend so that the friend could remind her of a name of a plant (Figure \ref{fig:P3_Message_Friend}). 

Participants appeared to take distinct roles in their online engagements, as either consumers or producers of information. Some participants (P1, P2, P5, P8) acknowledged they were more likely to be consumers of information than creators. Strikingly, these gardeners who self-identified as experienced in order to take part in the study, and taught us during participant observations, felt less comfortable sharing their knowledge online than they did in person. For example, P1 has been gardening for several years but when discussing helping identify plants on a website he frequents, he feels he \textit{"can't do that. Online, people know a lot more about [that]."} On the other hand, P7, who had been a master gardener, felt \textit{"people are going to learn from me instead of me learning from them."}

\subsubsection{Learning from what is left behind by others}

\begin{figure}[h!]
\centering
\includegraphics[width=\linewidth]{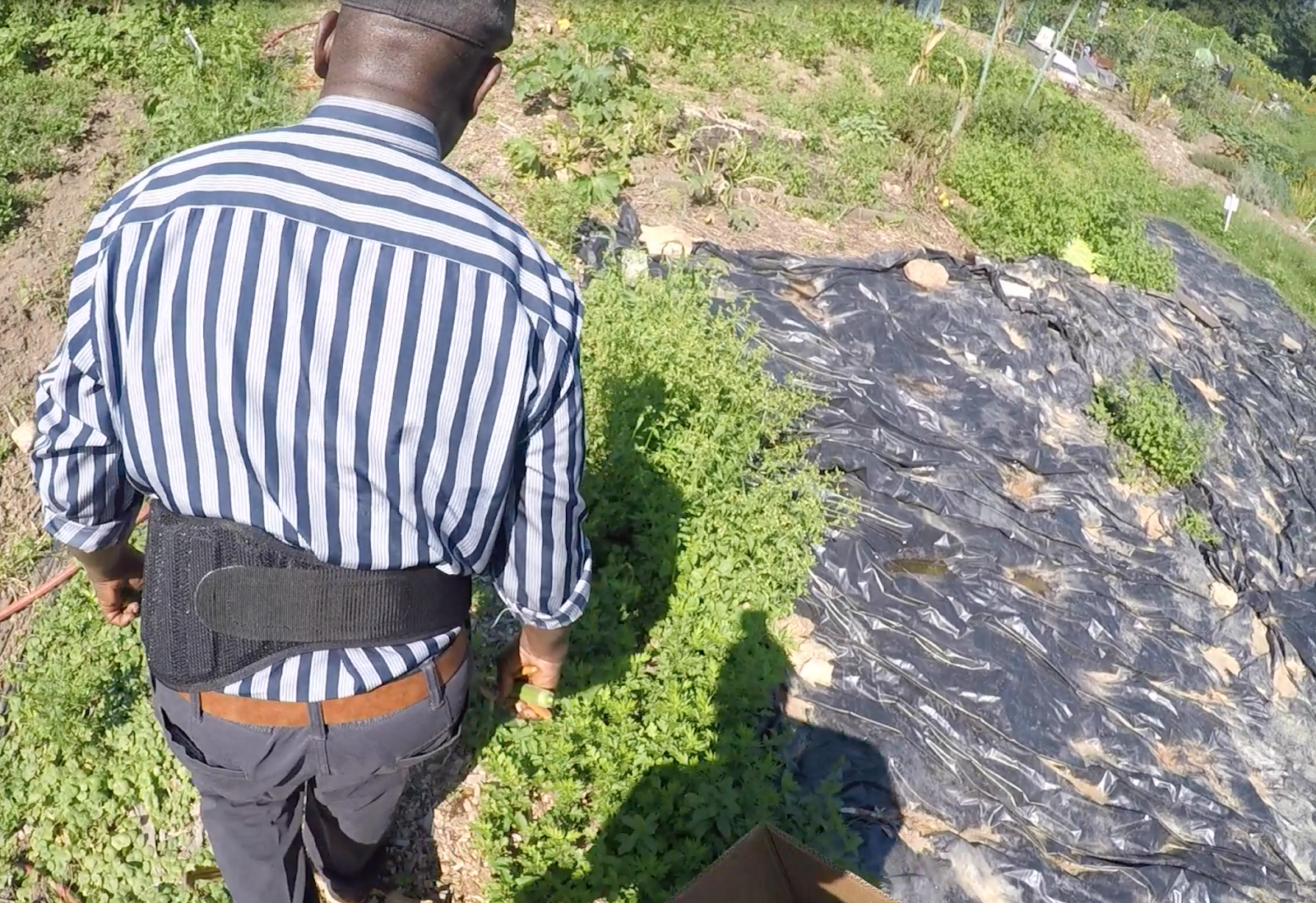}
\caption{P8 used plastic sheeting after observing its usage by farmers.}
\label{fig:P8_plastic}
\end{figure}

Gardeners also learn without the active involvement or even presence of others through examining the state of others' gardens and gardening configurations. Past work has found that gardeners learn from observing what farmers are doing, for example, from looking at the produce in farmers markets to understand what is in season \cite{Lyle:2015:GFC:2768545.2768549}. We found that gardeners can also learn techniques from observing farms. P8 described experimenting with a plastic sheet on the ground to prevent weed growth after observing this strategy in farms and searching online to understand the reasoning for it (See Figure \ref{fig:P8_plastic}).  

One participant observed others' gardens not just to learn from them, but to gauge how his own garden might be faring. P5 said that when on vacation, 
\textit{"I'll look at other people's plants to see, 'Okay, where are your tomatoes going now? What's the story here?' ... Because, if I'm in [state X] and we got a heat wave [in] both [state X] and [P5's home state Y], I'll kind of know what to expect when I get home."} By looking at similar plants in states with similar weather conditions, P5 is able to gauge the status of his own plant remotely. In these examples we see that it is not only the experienced practitioners such as farmers whose plots can reveal information and help relative novices make sense of why they do what they do, but also the gardens of peers, with which novices can use their own experience to compare their progress. 

\begin{figure}[h]
\centering
  \begin{subfigure}[b]{0.4\linewidth}
    \includegraphics[width=\linewidth]{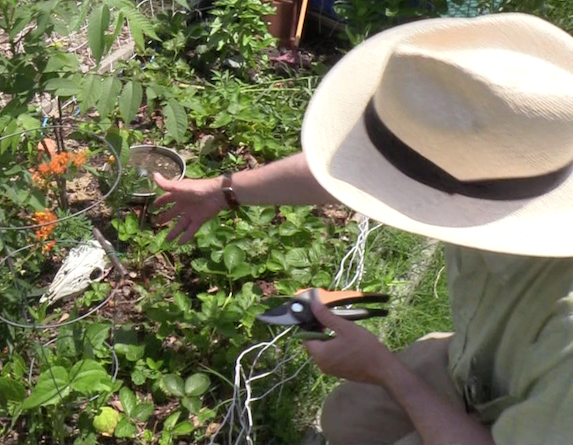}
    \includegraphics[width=\linewidth]{Images/PO3_antler1_2}
  \end{subfigure} \begin{subfigure}[b]{0.55\linewidth}
    \includegraphics[width=\linewidth]{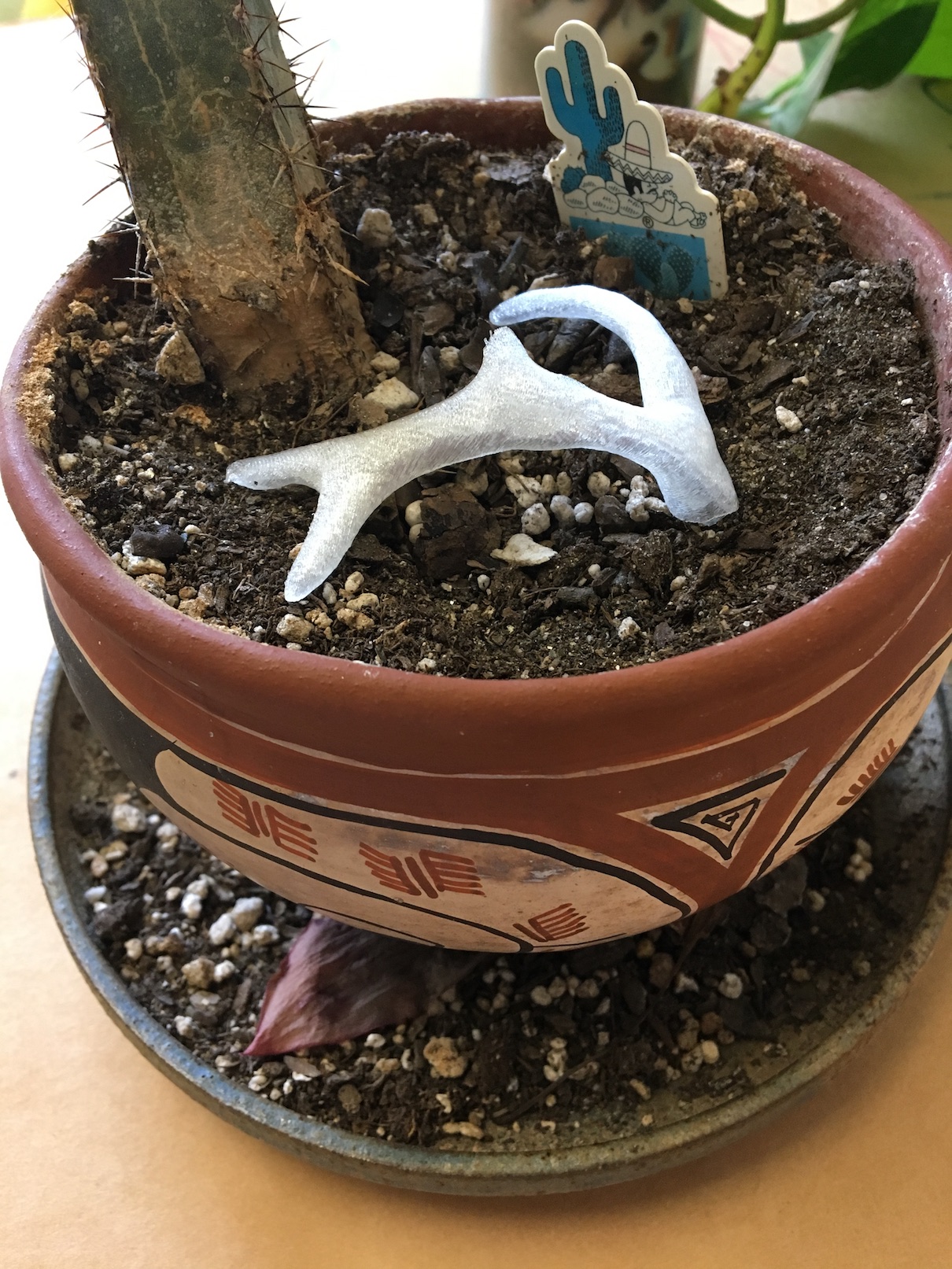}
  \end{subfigure}
  \caption{P1 pointing at deer skull (top left). P3 holding deer antlers (bottom left). Researcher's 3D printed antler (right).}
  \label{fig:P1_P3_antler}
\end{figure}

Learning new skills via observation of other gardens happened within our own research team during the course of the study. We observed that P1, P1's neighbor, and P3 all had deer antlers or bones lying around in their gardens to provide nourishment to the animals and the soil. During the writing of this paper, the second author found that the first author had added a small 3D printed set of antlers to his own indoor plant in the office (Figure \ref{fig:P1_P3_antler}). Reflecting on his experience, the first author saw this action as a novice imitating the experienced gardeners as a way to feel more connected to the community of gardeners with whom he had worked. This goal has been described in learning theory research as motivating and providing meaning to the process of becoming knowledgeably skillful in situated learning \cite{lave_wenger_1991}.

\section{Discussion}
Our findings reveal an understanding of sociality in the garden. We find that sociality exists in co-located and remote interactions with other gardeners, as well as with the broader community outside the garden. Practitioners’ sociality preferences are non-static and diverse. The garden space is designed purposefully in response to these sociality preferences to allow or restrict access to outsiders or other practitioners. Based on these findings, we present the following considerations for HCI in designing for collaboration in the garden. Specifically, we consider how the way sociality exists in the garden bears on  teaching or learning skills associated with the activity, social computing approaches in the garden space, and creating a sense of community inclusion. Given that past work urges designers to be mindful of the interactions between technology and practitioner sensibilities \cite{ Odom:2010:MWD:1858171.1858211, Lyle:2015:GFC:2768545.2768549}, we also present potential tensions that arise in introducing technology to the garden in each section. 
\subsection{Teaching and learning embodied skills in the garden} 
Promoting lifelong learning is described as one of the "Grand Challenges" for HCI \cite{Shneiderman:2016:GCH:2991131.2977645}. Our work highlights skill sharing as a key domain for design in the garden. Teaching and learning are regular practices of the participants we studied. Skills are continuously gained and refined, both through direct interaction with others as well as observations of others' gardens. Yet many individuals lack opportunities for in-person learning, and a lack of gardening knowledge and access to experts has been linked to programs to foster gardening failing \cite{Davis2017}. Below, we describe opportunities to leverage the expertise of experienced gardeners as a way to support multigenerational interaction and cultural exchange through observation and practise. 

\textit{Learning by doing}, under the instruction of expert family members and friends, was an important way that many participants in our study -- as well as the first author as a participant observer -- gained initial gardening skills. Though participants spoke of using platforms such as Facebook to look up gardening questions or sending a picture to their friends on the phone, these forms of media are not sufficient for learning many of the embodied skills key to gardening. In other settings, research has examined ways to support embodied learning through skill demonstration, when an expert and novice are not co-located. Future work can draw on past work on embodied learning to encourage interaction with sensory stimuli that mimic an expert practitioner, for example, imitating a projected video of an expert \cite{Xiao:2016:IEI:2858036.2858577} or experiencing vibrations synchronized to the movement of an artisan using a tool \cite{JapaneseArtisan2013}. 

 There are also open opportunities to create remote real-time skill sharing for outdoor activities such as gardening that draws on the telepresence literature, which includes research o has other concepts in the area of real-time interaction between distributed groups of practitioners such as ego-centric feedback \cite{Han:2016:AAV:2875194.2875237} and tangible interaction with remote physical objects \cite{Leithinger:2014:PTS:2642918.2647377}. When considering harnessing expert experience of gardeners, however, it is important to note that while some experienced practitioners, such as participant P7, might feel confident that gardeners will learn from them, others, such as P1, might be less inclined to share their knowledge. Taking the initiative to share information, rather than primarily being an information consumer, may depend on whether the practitioner feels that other people in the group “know a lot more,” as P1 put it. This self-perception of the practitioners’ expertise relative to the group is an important consideration when encouraging knowledge-sharing behaviors. Some may be far more willing to share one-to-one than in larger groups. 

Our study adds to past work on how gardeners learn from direct interaction with other gardeners \cite{rahm2002emergent}. We find that learning can also take place through making sense of the traces left by other gardeners, with some gardeners mimicking techniques that they learned from their observations. The traces that participants focused on often had to do with gaining an awareness of how more experienced gardeners might approach sustainability. For example, P8’s technique of plastic sheeting was based on how farmers discouraged weeds without chemicals. P3 and P1’s ideas of placing antlers in the garden came from an understanding of the necessity to nourish the soil and other animals. Given that HCI research on gardening is often motivated by sustainability (e.g., \cite{Heitlinger:2013:SHG:2541016.2541023, Lyle:2015:GFC:2768545.2768549}), one design opportunity in this direction is to preserve and share the traces of skilled gardeners' routine actions.

\subsection{Designing for varying social preferences}
Participants created spaces that reflected their sociality preferences using different kinds of boundaries as a way to manage interactions with other gardeners and people outside the garden. Below, we describe the implications of these findings for how social technologies in the garden might be received.

In considering social technologies in the garden, it is essential to consider the ways that preferences for sociality are not constant. In our data, gardeners' preferences varied with changing moods, tasks at hand, and the constraints and possibilities of a particular gardening space. In accounting for varying sociality preferences in the design of collaborative systems, researchers have explored concepts like interruptibility and signaling availability primarily in the context of indoor work settings \cite{Fogarty:2005:ETE:1054972.1055018, Zuger:2017:RIW:3025453.3025662}. As one example, researchers have studied whether interruptibility can be estimated from whether productivity is affected by someone typing on a keyboard or standing with one or more guests in the vicinity \cite{Fogarty:2005:ETE:1054972.1055018}. In designing or modifying technologies to be context-aware in outdoor, recreational, and educational settings like in the garden, how might we translate this concept of interruptibility? Attending to the location of a gardener and the meaning that they assign to different locations is a first step in estimating willingness to be approached. Our findings reveal some areas in the garden are assigned significance based on the kinds of activities that take place, for example, a meditation hut would imply leaning towards solitude, whereas placing chairs together encourages interaction. Further, different configurations indicate openness to interacting with different kinds of audiences, and whether activities take place inside or outside the garden fence has meaning. P2 arranging chairs to create a more intimate space for people inside the garden is intended for close communication with loved ones, whereas the benches outside the fence of P8’s garden invites unknown outsiders to sit and observe. Gardeners’ willingness to interact socially or use technology at all might be estimated from their locations within the garden and the configurations that they create over time and in the moment. 

Even as we provide implications to avoid introducing technology in the garden due to its intrusiveness for some, others integrate certain types of technology into the gardening experience \cite{Goodman:2011:GGN:1978942.1979273}. Expanding on past studies that find gardeners using technology to coordinate with garden members and showcasing their ongoing activities (e.g. sharing photographs on blogs, social media) \cite{Goodman:2011:GGN:1978942.1979273, Heitlinger:2013:SHG:2541016.2541023}, we find that gardeners also use technology to encourage involvement in the garden. For example, participants shared volunteering opportunities and recipes online. And, overall, gardeners reported positive interactions sharing garden-related content (and particularly appreciated receiving compliments). We see the sharing of gardening-related content as one way to support a sense of community and civic engagement. Here, further research might consider how sociality varies across different activities or types of interactions and how this relates to activities seen as social or purposeful, for learning or for community, and by different kinds of practitioners. 


\subsection{Negotiating inclusion and ownership} 
Past work has noted that inclusion is a core value of community gardeners \cite{Heitlinger:2013:SHG:2541016.2541023,Wang:2015:ISS:2768545.2768556}. Our findings also reveal a desire for inclusion -- demonstrated through a physical configuration of gardens, from the use of chicken wire to the clearing of large branches -- is carefully balanced with gardeners' varying preferences for sociality. This tension between the access to skills experts can provide to novices and how novices can be excluded due to expert "territoriality" \cite{Thom-Santelli:2010:YKE:1753326.1753578} is evident here: with the territory applying quite literally to the physical spaces of gardens.  Below we discuss insights from our findings on how technology might affect the delicate balance between inclusion and ownership. 

From our findings, we see opportunities where outsiders could come to interact in garden spaces to, for example, learn sustainable behaviors or create a sense of community by complimenting growers.
Researchers can examine approaches that allow audiences to interact physically with gardening sites, such as location-based exploration concepts such as geocaching \cite{O'Hara:2008:UGP:1357054.1357239} or even citizen science approaches that have people make data about physical spaces such as backyards, local parks, and other environmental observations accessible to the general public (e.g., Phenology Maps \cite{usnpa}, NatureNet \cite{preece2016crowdsourcing}, and YardMap \cite{Kobori2016}). These approaches resonate with aims in the gardening space, such as learning when a specific plant species will bloom \cite{usnpa} or the environmental impact of personal growing practices \cite{Dickinson2014}. When designing for public digitally-mediated interaction in the gardening space, however, it is necessary to think about how one might encourage the community to respect boundaries established by the inhabitants of the gardening space. One approach might support gardeners in indicating that the local community is welcome to interact with certain elements or parts of the garden space (e.g., P1's flags for Independence Day) or inviting volunteers for particular tasks (e.g., P6’s volunteers helping with weeding and harvesting produce). The metaphors of different kinds of fences and boundaries to promote or restrict visibility and access can inspire design in this area.


In our study, gardeners expressed a sense of ownership with the gardening space and the plants that they cultivate by establishing physical boundaries. Our findings also show examples of participants cultivating relations through their activities inside the gardening space around their native plants and other memory objects. What does it mean to design for inclusiveness in a living space whose inhabitants feel responsible for, and when the objects inside the space hold meaning for the people or communities close to them? We propose that there are opportunities in HCI to support gardeners in highlighting and sharing the meaning that different objects or plants hold for them and their community. A current project that might be seen as falling in this design space is the Connected Seeds project that attempts to connect people to their heritage through food by collecting and sharing stories related to locally-grown seeds \cite{Heitlinger:2018:CSS:3210604.3210620}. Further areas for connection we identify from our work include the concept of memory objects that reflect the gardener's relationship with a loved one, learning techniques for sustainable growing through observation, and experiencing different cultures (e.g., sharing produce native to the gardener’s country). 

An important aspect of inclusion is ensuring access for people with disabilities and mobility constraints -- a priority mentioned by gardeners in our study. A fruitful future direction is to investigate technology's role in supporting accessible gardening. Research has in the past explored approaches to bringing the experience of a remote location through an on-site physical proxy to a user. For example, the Telegarden uses a robotic arm to interact with a remote shared garden such as in \cite{Kahn:2005:ROMAN:1513749}, and the Teletourism system provides accessible tourism experiences through video chat with a video-sharer at the actual physical location that the viewer would like to experience \cite{deGreef:2016:TIT:2818052.2869082}. This approach of virtually visiting a space might not be appealing when trying to communicate tacit knowledge that requires certain sensory stimuli (e.g., learning to determine the ripeness of produce via touch). Further, we propose that in addition to focusing on enjoyment, engagement, or immersion, as these prior systems do, it is important to think about how design in this space can position people and the kind of connections it can enable. For example, volunteers of different expertise levels connected with experienced gardeners P6, P7, and P9 in a community learning garden, where they worked with and learned from each other while contributing to the community’s food security by donating produce. In other words, the garden is not just a space for recreation and connecting with nature – it is also a meeting point for practitioners that provides opportunities to be good citizens. One area for future research might involve supporting experts who are no longer able to garden in remotely sharing valuable skills with novices. This kind of approach could provide much needed experience, a lack of which has posed challenges to previous projects \cite{Davis2017}. In proposing this idea, we do not intend to minimize the real need to assess and improve the accessibility of outdoor spaces, a topic addressed in past work through crowdsourcing \cite{Saha:2019:PSW:3290605.3300292}.

\section{Conclusion}
This paper contributes an understanding of how sociality is configured in the garden environment, and the ways that skill sharing take place. Through our participant observations, we found that gardeners use and modify the boundaries of their gardens to maintain a balance between two considerations: need for a personal space for themselves, their co-gardeners, or loved ones, and a motivation to create an inclusive space in and around the garden and show community sentiment. Social skill sharing occurs between gardeners with a focus on on-site interactions that lead to learning. In addition to direct learning interactions through observation, indirect modes of learning take place via observation of other people's gardens. We contribute a discussion of design considerations to support interactions for skill sharing between users with different expertise levels, to support varying sociality preferences, and in negotiating the tensions between community inclusion and ownership in the garden nature-space.

\section{Acknowledgements}
We thank those who participated in this study for sharing their perspectives and allowing us to learn from them in their gardens.
We are also grateful to those who have reviewed versions of this paper, including Wayne G. Lutters, Catherine Plaisant, Norman Makoto Su, Samuel Philip Goree, Majdah Alshehri, Leslie S. Liu, and anonymous reviewers. 
\balance{}

\bibliographystyle{SIGCHI-Reference-Format}
\bibliography{proceedings}

\end{document}